\documentclass[12pt]{article}
\usepackage{graphicx}
\usepackage{amssymb,amsmath}
\usepackage{cite}

\newcommand{\re}{\mathop{\mathrm{Re}}}

\begin{document}

\title{Modeling of flows with the power-law spectral densities and power-law distributions of flow's intensities}

\author{Bronislovas Kaulakys,
Miglius Alaburda,
Vygintas Gontis,\\
Tadas Meskauskas and
Julius Ruseckas\\
Institute of Theoretical Physics and Astronomy of Vilnius University, \\
A. Gostauto 12, LT-01108 Vilnius, Lithuania
}

\date{}

\maketitle 

\begin{abstract}
We present analytical and numerical results of modeling of flows represented as the correlated non-Poissonian point process and as  the Poissonian sequence of pulses of the different size. Both models may generate signals with the power-law distributions of the intensity of the flow and the power-law spectral density. Furthermore, different distributions of the interevent time of the point process and different statistics of the size of pulses  may result in $ 1/f^{\beta} $ noise with $ 0.5\lesssim\beta\lesssim2 $. Combination of the models is applied for modeling of the Internet traffic. 
\end{abstract}

\section{Introduction}

Modeling and simulations enable one to understand and explain the observable phenomena and predict the new ones. 
This is true, as well, for the mathematical study and modeling of the traffic flow with the aim to get a better understanding of phenomena and avoid some problems of traffic congestion. Traffic phenomena are complex and nonlinear, they show cluster formation, huge fluctuations and long-range dependencies. Almost forty years ago from the empirical data it was detected that fluctuations of a traffic current on a expressway obey the $1/f$ law for low spectral frequencies \cite{musha}. Similarly $1/f$ noise is observable in the flows of granular materials \cite{schick,peng}. 

$1/f$ noise, or $1/f$ fluctuations are usually related with the power-law distributions of other statistics of the 
fluctuating signals, first of all with the power-law decay of autocorrelations and the long-memory processes (see, 
e.g., comprehensive bibliography of $1/f$ noise in the website \cite{li}, review articles \cite{weismann,wong} and 
references in the recent paper \cite{KG}). The appearance of the clustering and large fluctuations in the traffic 
and granular flows may be a result of synchronization of the nonlinear systems with stopping and driven by common noise, resulting in the nonchaotic behavior of the Brownian-type motions, intermittency and $1/f$ noise \cite{KV,KI}. 

The traffic and granular flows usually may be considered as those consisting of the discrete identical objects such as vehicles, pedestrians, granules, packets and so on, they may be represented as consisting of pulses or elementary events and further simplified to the point process model \cite{KG,zhang,KM,K}. Moreover, from the modeling of the traffic it was found that $1/f$ noise may be the result of clustering and jumping \cite{zhang} similar to the point process model of $1/f$ noise \cite{KG,KM,K}. 

On the other hand, $1/f$ noise may be conditioned by the flow consisting of uncorrelated pulses of variable size with the power-law distribution of the pulse durations \cite{RK}. In the Internet traffic the flow of the signals primarily is composed of the power-law distributed file sizes. The files are divided by the network protocol into the equal packets \cite{field}. Therefore, the total incoming web traffic is a sequence of the packets arising from large 
number of requests. Such a flow exhibits $1/f$ fluctuations, as well \cite{field,GKR}. 

The long-range correlations and the power-law fluctuations in the wide range of the time scale from minutes to months 
of the expressway traffic flow have recently been observed and investigated using the method of the detrended fluctuation analysis \cite{tadaki}. There are no explanations why the traffic flow exhibit $1/f$ noise behavior in such a large interval of the time. 

It is the purpose of this paper to present analytical and numerical results for the modeling of flows represented as 
sequences of different pulses and as a correlated non-Poissonian point process resulting in $1/f$ noise and to apply these results to the modeling of the Internet traffic.

\section{Signal as a sequence of pulses}

We will investigate a signal of flow consisting of a sequence of pulses, 
\begin{equation}
I(t)=\sum_{k}A_{k}(t-t_{k}).  
\label{sig:1}
\end{equation}
Here the function $A_{k}(t-t_{k})$ represents the shape of the $k$ pulse having influence on the signal $I(t)$ in the 
region of time $t_{k}$.

\subsection{Power spectral density}

The power spectral density of the signal \eqref{sig:1} can be written as 
\begin{equation}
S(f)=\lim_{T\to\infty}\left\langle\frac{2}{T}\sum_{k,k^{\prime}}e^{i\omega(t_{k}-t_{k^{\prime}})}
\int\limits_{t_{i}-t_{k}}^{t_{f}-t_{k}}
\int\limits_{t_{i}-t_{k^{\prime}}}^{t_{f}-t_{k^{\prime}}}
A_{k}(u)A_{k^{\prime}}(u^{\prime})e^{i\omega(u-u^{\prime})}dudu^{\prime}\right\rangle,
\label{psd:1}
\end{equation}
where $\omega=2\pi f$, $T=t_{f}-t_{i}\gg\omega^{-1}$ is the observation time and the brackets $\langle\ldots\rangle$ 
denote the averaging over realizations of the process.
We assume that pulse shape functions $A_{k}(u)$ decrease sufficiently fast when $|u|\to\infty$. Since $T\to\infty$, 
the bounds of the integration in Eq.~\eqref{psd:1} can be changed to $\pm\infty$.
 
When the time moments $t_{k}$ are not correlated with the shape of the pulse $A_{k}$, the power spectrum is \cite{schick}
\begin{equation}
S(f)=\lim_{T\to\infty}\frac{2}{T}\sum_{k,k^{\prime}}\left\langle e^{i\omega(t_{k}-t_{k^{\prime}})}\right\rangle
\left\langle\int\limits_{-\infty}^{+\infty}\int\limits_{-\infty}^{+\infty}
A_{k}(u)A_{k^{\prime}}(u^{\prime})e^{i\omega(u-u^{\prime})}dudu^{\prime}\right\rangle.
\label{psd:2}
\end{equation}

After introduction of the functions \cite{RK} 
\begin{equation}
\Psi_{k,k^{\prime}}(\omega)=\left\langle\int\limits_{-\infty}^{+\infty}A_{k}(u)e^{i\omega u}du
\int\limits_{-\infty}^{+\infty}A_{k^{\prime}}(u^{\prime})e^{-i\omega u^{\prime}}du^{\prime}\right\rangle
\label{psd:3}
\end{equation}
and
\begin{equation}
\chi_{k,k^{\prime}}(\omega)=\left\langle e^{i\omega(t_{k}-t_{k^{\prime}})}\right\rangle
\label{psd:4}
\end{equation}
the spectrum can be written as
\begin{equation}
S(f)=\lim_{T\to\infty}\frac{2}{T}\sum_{k,k^{\prime}}\chi_{k,k^{\prime}}(\omega)\Psi_{k,k^{\prime}}(\omega).
\label{psd:5}
\end{equation}

\subsection{Stationary process}

Equation \eqref{psd:5} can be further simplified for the stationary process. 
Then all averages can depend only on $k-k^{\prime}$, i.e., 
\begin{equation}
\Psi_{k,k^{\prime}}(\omega)\equiv\Psi_{k-k^{\prime}}(\omega)
\label{sp:1}
\end{equation}
and
\begin{equation}
\chi_{k,k^{\prime}}(\omega)\equiv\chi_{k-k^{\prime}}(\omega).
\label{sps:2}
\end{equation}
Equation \eqref{psd:5} then reads
\begin{equation}
S(f)=\lim_{T\to\infty}\frac{2}{T}\sum_{k,k^{\prime}}\chi_{k-k^{\prime}}(\omega)\Psi_{k-k^{\prime}}(\omega).
\label{sps:3}
\end{equation}
Introducing a new variable $q\equiv k-k^{\prime}$ and changing the order of summation, yield 
\begin{multline}
S(\omega)=\lim_{T\to\infty}\frac{2}{T}\sum_{q=1}^{k_{\max}-k_{\min}}
\sum_{k=k_{\min}}^{k_{\max}-q}\chi_{q}(\omega)\Psi_{q}(\omega)\\
+\lim_{T\to\infty}\frac{2}{T}\sum_{q=k_{\min}-k_{\max}}^{-1}
\sum_{k=k_{\min}-q}^{k_{\max}}\chi_{q}(\omega)\Psi_{q}(\omega)
+\lim_{T\to\infty}\frac{2}{T}\sum_{k=k_{\min}}^{k_{\max}}\Psi_{0}(\omega).
\label{sp:4}
\end{multline}
Here $k_{\min}$ and $k_{\max}$ are minimal and maximal values of the index $k$ in the interval of observation $T$. 
Eq.~\eqref{sp:4} may be simplified to the structure 
\begin{equation}
S(f)=2\bar{\nu}\Psi_{0}(\omega)+\lim_{T\to\infty}4\sum_{q=1}^{N}\left(\bar{\nu}-\frac{q}{T}\right)
\re\chi_{q}(\omega)\Psi_{q}(\omega)
\label{sp:5}
\end{equation}
where $\bar{\nu}$ is the mean number of pulses per unit time and $N=k_{\max}-k_{\min}$ is the number of pulses in the time interval $T$. 

If the sum $\frac{1}{T}\sum_{q=1}^{N}q\re\chi_{q}(\omega)\Psi_{q}(\omega)\to 0$ when $T\to\infty$, then the second term 
in the sum vanishes and the spectrum is
\begin{equation}
S(f)=2\bar{\nu}\Psi_{0}(\omega)+4\bar{\nu}\sum_{q=1}^{\infty}\re\chi_{q}(\omega)\Psi_{q}(\omega)
=2\bar{\nu}\sum_{q=-\infty}^{\infty}\chi_{q}(\omega)\Psi_{q}(\omega).
\label{sp:7}
\end{equation}

\subsection{Fixed shape pulses}

When the shape of the pulses is fixed ($k$-independent) then the function $\Psi_{k,k^{\prime}}(\omega)$ does not 
depend on $k$ and $k^{\prime}$ and, therefore, $\Psi_{k,k^{\prime}}(\omega)=\Psi_{0,0}(\omega)$. 
Then equation \eqref{psd:5} yields the power spectrum
\begin{equation}
S(f)=\Psi_{0,0}(\omega)\lim_{T\to\infty}\frac{2}{T}\sum_{k,k^{\prime}}\chi_{k,k^{\prime}}(\omega)
\equiv\Psi_{0,0}(\omega)S_{\delta}(\omega).
\label{fsp:1}
\end{equation}
Eq.~\eqref{fsp:1} represents the spectrum of the process as a composition of the spectrum of one pulse, 
\begin{equation}
\Psi_{0,0}=\left|\int\limits_{-\infty}^{+\infty}A_{k}(t)e^{i\omega t}dt\right|^{2},
\label{fsp:2}
\end{equation}
and the power density spectrum $S_{\delta}(\omega)$ of the point process 
\begin{equation}
I_{\delta}(t)=a\sum_{k}\delta(t-t_{k})
\label{fsp:3}
\end{equation}
with  the area of the pulse $a=1$.

\section{Stochastic point processes}

The shapes of the pulses mainly influence the high frequency power spectral density, i.e., at $\omega\ge 1/\Delta t_{p}$, with $\Delta t_{p}$ being the characteristic pulse length. Therefore the power spectral density at low frequencies for the not very long pulses is mainly conditioned by the correlations between the transit times $t_{k}$, i.e., the signal may be approximated by the point process. 

The point process model of $1/f^{\beta}$ noise has been proposed \cite{KM,K}, generalized \cite{KG}, analysed and used 
for the financial systems \cite{GK}. It has been shown that when the average interpulse, interevent, interarrival, 
recurrence or waiting times $\tau_{k}=t_{k+1}-t_{k}$ of the signal diffuse in some interval, the power spectrum of such 
process may exhibit the power-law dependence, $S_{\delta}(f)\sim 1/f^{\beta}$, with $0.5\lesssim\beta\lesssim 2$. 
The distribution density of the signal \eqref{fsp:3} intensity defined as $I=1/\tau_{k}$ may be of the power-law, $P(I)\sim I^{-\lambda}$, with $2\leqslant\lambda\leqslant 4$, as well. The exponents $\beta$ and $\lambda$ are depending on the manner of diffusion-like motion of the interevent time $\tau_{k}$ and, e.g., for the multiplicative process are interrelated \cite{KG,GK}. For the pure multiplicative process \cite{KG} 
\begin{equation}
\beta=1+\alpha,\quad \lambda=3+\alpha,
\label{spp:1}
\end{equation}
where $\alpha$ is the exponent of the power-law distribution, $P_{k}(\tau_{k})\sim\tau_{k}^{\alpha}$, of the interevent 
time. In general, for relatively slow fluctuations of $\tau_{k}$, the distribution density of the flow $I$, 
\begin{equation}
P(I)\sim P_{k}(I^{-1})I^{-3},
\label{spp:2}
\end{equation}
is mostly conditioned by the multiplier $I^{-3}$. As far as the point process model has recently \cite{KG,GK} been analysed rather properly here we will not repeat the analysis and only present some new illustrations. 

\begin{figure}[tbh]
\begin{center}
\includegraphics[width=.5\textwidth]{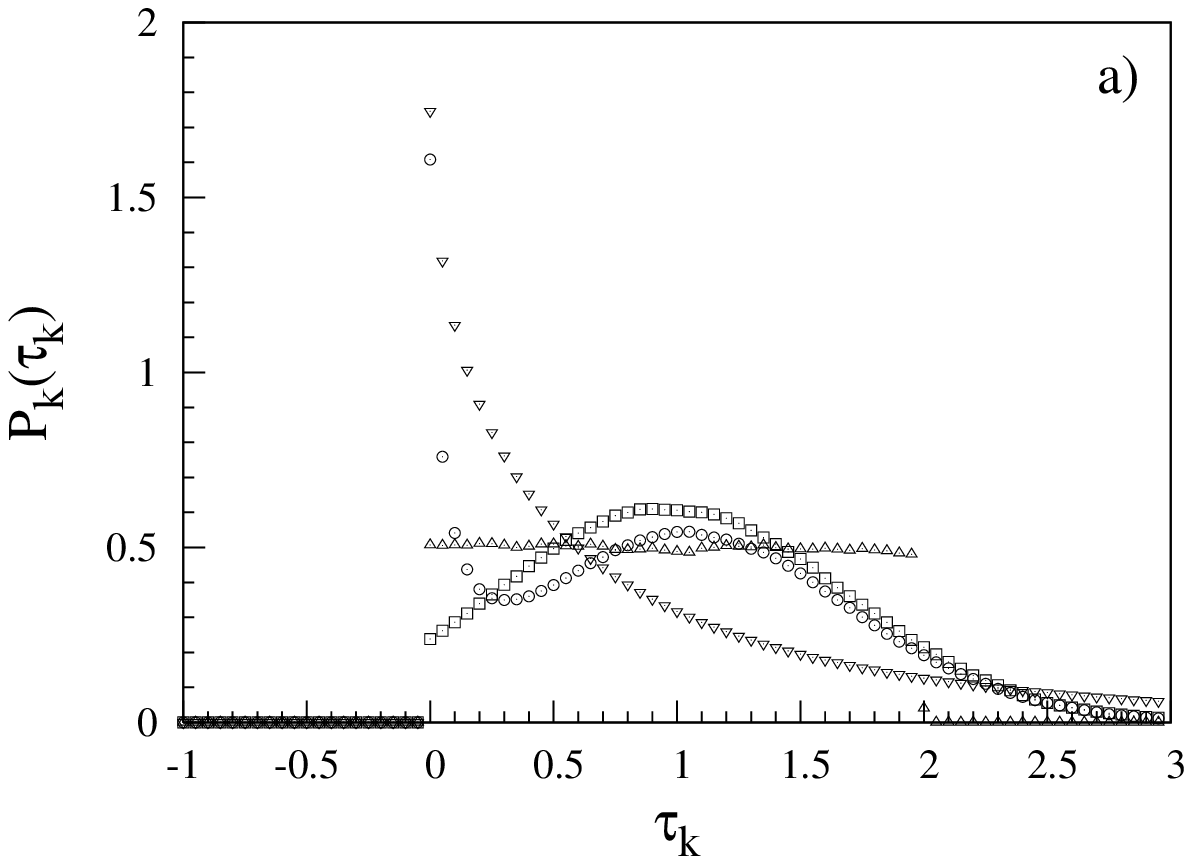}
\includegraphics[width=.5\textwidth]{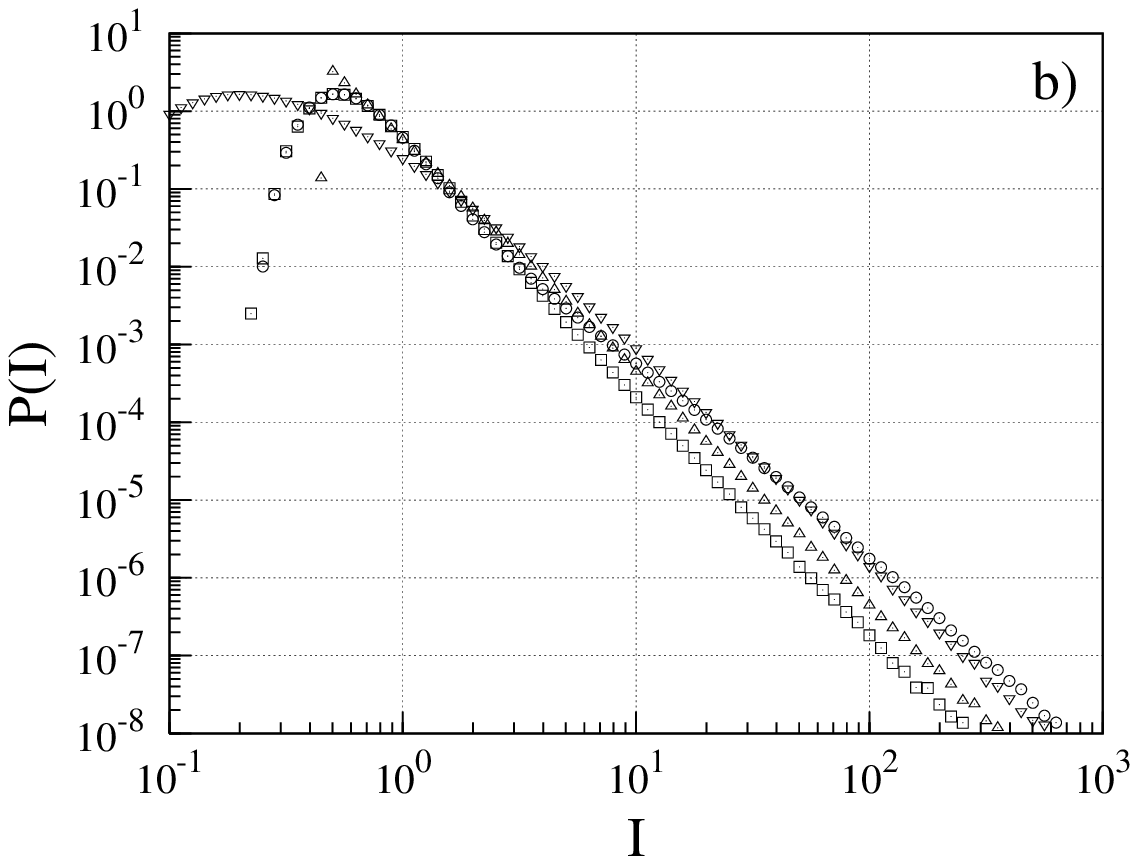}
\hspace{-10pt}
\includegraphics[width=.5\textwidth]{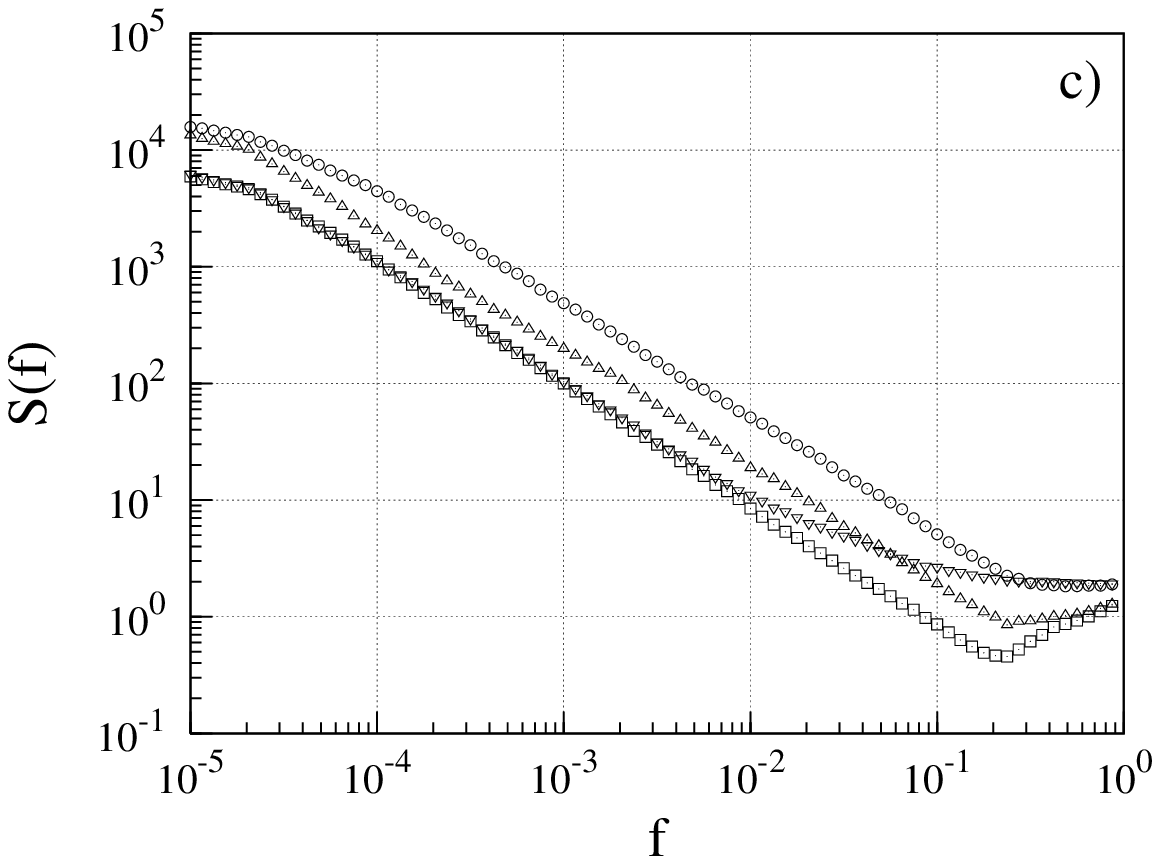}
\end{center}
\vspace{-10pt}
\caption{Distribution densities of the interevent time $\tau_{k}$, (a), of the flow $I(t)$, (b), and of the power spectra $ S(f) $, (c), for different point processes with the slow diffusion-like motion of the average interevent time. Different symbols correspond to different types of the generation of the interevent sequences.}
\label{spp:dist}
\end{figure}
Figure \ref{spp:dist} demonstrates that for essentially different distributions of $\tau_{k}$, the power spectra and 
distribution densities of the point processes are similar. 

Further we proceed to the flow consisting of the pulses of different durations and application of this approach for 
modeling of the Internet traffic.

\section{Flow consisting of pulses of variable duration}
\label{sec:flow}

When the occurrence times $t_{k}$ of the pulses are uncorrelated and distributed according to the Poisson process, the 
power spectrum of the random pulse train is given by the Carlson's theorem
\begin{equation}
S(f)=2\bar{\nu}\left\langle |F_{k}(\omega)|^{2}\right\rangle,
\label{fcf:1}
\end{equation}
where
\begin{equation}
F_{k}(\omega)=\int\limits_{-\infty}^{+\infty}A_{k}(t)e^{i\omega t}dt
\label{fcf:2}
\end{equation}
is the Fourier transform of the pulse $A_{k}$. Suppose that the random parameters of the pulses are the duration and the area (integral) of the pulse. We can take the form of the pulses as 
\begin{equation}
A_{k}(t-t_{k})=T_{k}^{\rho}A\left(\frac{t-t_{k}}{T_{k}}\right),
\label{fcf:3}
\end{equation}
where $T_{k}$ is the characteristic duration of the pulse. The value of the exponent $\rho=0$ corresponds to the fixed 
height but different durations, the telegraph-like pulses, whereas $\rho=-1$ corresponds to constant area pulses but of different heights and durations, and so on. 

For the power-law distribution of the pulse durations,
\begin{equation}
P(T_{k})=
\begin{cases}
\frac{\delta+1}{T_{\max}^{\delta+1}-T_{\min}^{\delta+1}}T_{k}^{\delta}, 
& T_{\min}\leq T_{k}\leq T_{\max},\\
0, &\mbox{otherwise},
\end{cases}
\label{fcf:4}
\end{equation}
from Eqs.~\eqref{fcf:1} and \eqref{fcf:2} we have the spectrum
\begin{equation}
S(f)=\frac{2\bar{\nu}(\delta+1)}{(T_{\max}^{\delta+1}-T_{\min}^{\delta+1})\omega^{\delta+2\rho+3}}
\int\limits_{\omega T_{\min}}^{\omega T_{\max}}|F(u)|^{2}u^{\delta+2\rho+2}du.
\label{fcf:5}
\end{equation}
For $\tau_{\max}^{-1}\ll\omega\ll\tau_{\min}^{-1}$ when $\delta>-1$ the expression \eqref{fcf:5} may be approximated as 
\begin{equation}
S(f)\approx\frac{2\bar{\nu}(\delta+1)}{(T_{\max}^{\delta+1}-T_{\min}^{\delta+1})\omega^{\delta+2\rho+3}}
\int\limits_{0}^{\infty}|F(u)|^{2}u^{\delta+2\rho+2}du.
\label{fcf:6}
\end{equation}
Therefore, the random pulses with the appropriate distribution of the pulse duration (and area) may generate signals 
with the power-law distribution of the spectrum with different slopes. So, the pure $1/f$ noise generates, e.g., the fixed area ($\rho=-1$) with the uniform distribution of the durations ($\delta=0$) sequences of pulses, the fixed height ($\rho=0$) with the uniform distribution of the inverse durations $\gamma=T_{k}^{-1}$ and all other sequences of random pulses satisfying the condition $\delta+2\rho=-2$.

In such a case from Eq.~\eqref{fcf:6} we have
\begin{equation}
S(f)\sim\frac{(\delta+1)\bar{\nu}}{(T_{\max}^{\delta+1}
-T_{\min}^{\delta+1})f}.
\label{fcf:7}
\end{equation}

\section{The Internet traffic}

In this Section we will apply the results of the Section \ref{sec:flow} for modeling the Internet traffic. The 
incoming traffic consists of sequence of packets, which are the result of the division of the requested files by 
the network (TCP) protocol. Maximum size of the packet is $1500$ bytes. Therefore, the information signal is as the point 
process \eqref{fsp:3} with the pulse area $a=1500$ bytes. Further, we will analyse the flow of the packets and will 
measure the intensity of the flow in packets per second. In such a system of units in Eq.~\eqref{fsp:3} we should put 
$a=1$.

\begin{figure}[tbh]
\begin{center}
\includegraphics[width=.5\textwidth]{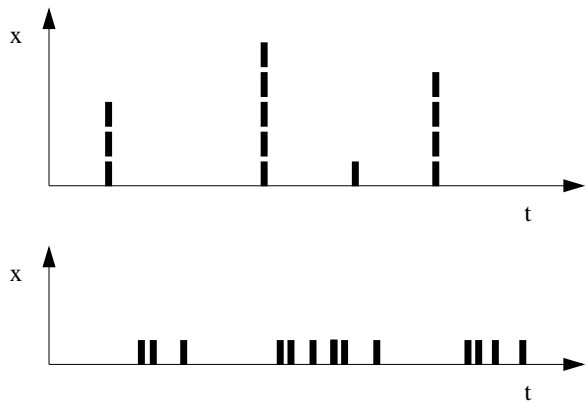}
\end{center}
\vspace{-10pt}
\caption{Division of the requested files into equal size packets with some inter-packet time.}
\label{tit:packets}
\end{figure}

We exploit the empirical observation \cite{field,field2} that the distribution of the file sizes $x$ may be 
described by the positive Cauchy distribution
\begin{equation}
P(x)=\frac{2}{\pi}\frac{s}{s+x^{2}}
\label{tit:1}
\end{equation}
with the empirical parameter $s=4100$ bytes. This distribution asymptotically exhibits the Pareto distribution and follows the Zipf's law $ P\left( X >x\right) \sim1/x$. The files are divided by the network protocol into packets of the maximum size of $1500$ bytes or less. In the Internet traffic the packets spread into the Poissonian sequence with the average inter-packet time $\tau_{p}$ (see Fig.~\ref{tit:packets}). The total incoming flow of the packets to the server consists of the packets arising from the Poissonian request of the files with the average interarrival time of files $\tau_{f}$. 

\begin{figure}[tbh]
\begin{center}
\includegraphics[width=.5\textwidth]{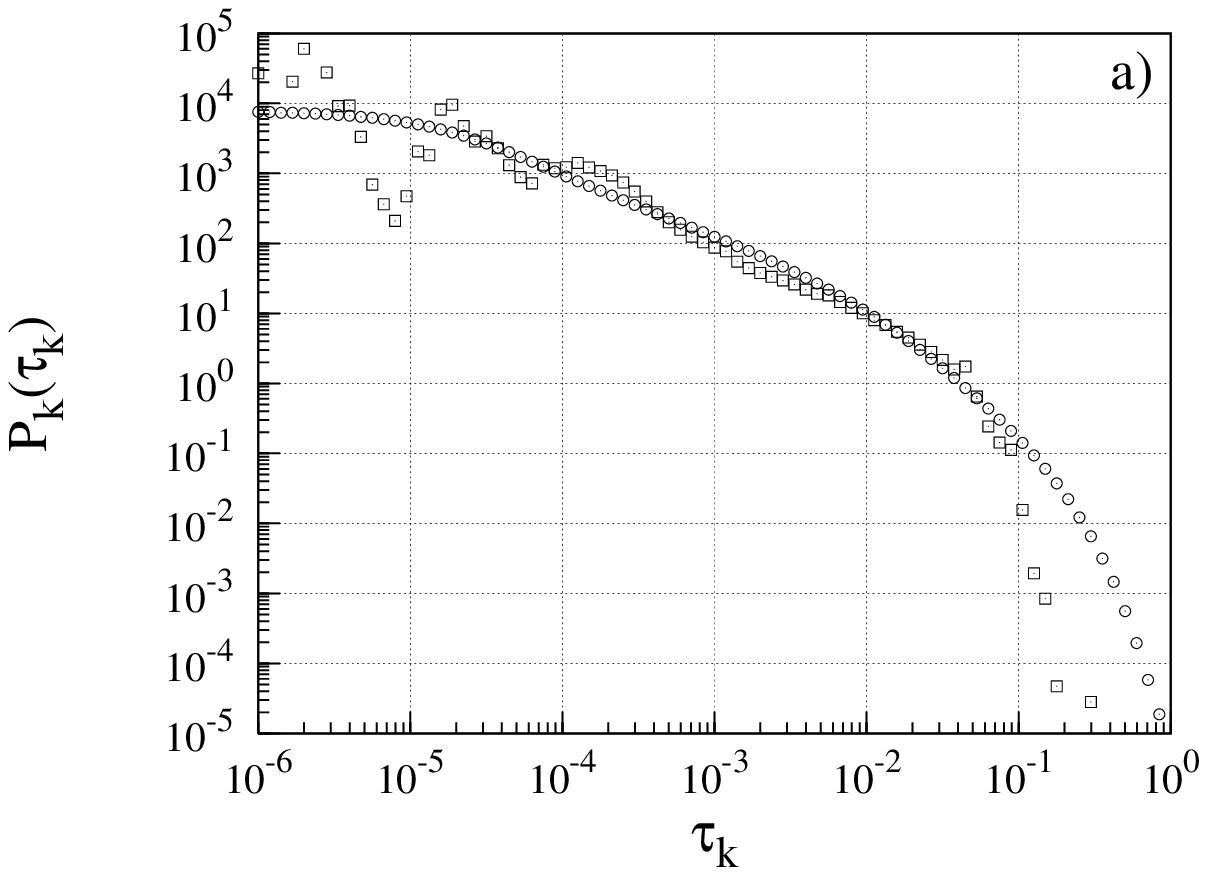}
\hspace{-10pt}
\includegraphics[width=.5\textwidth]{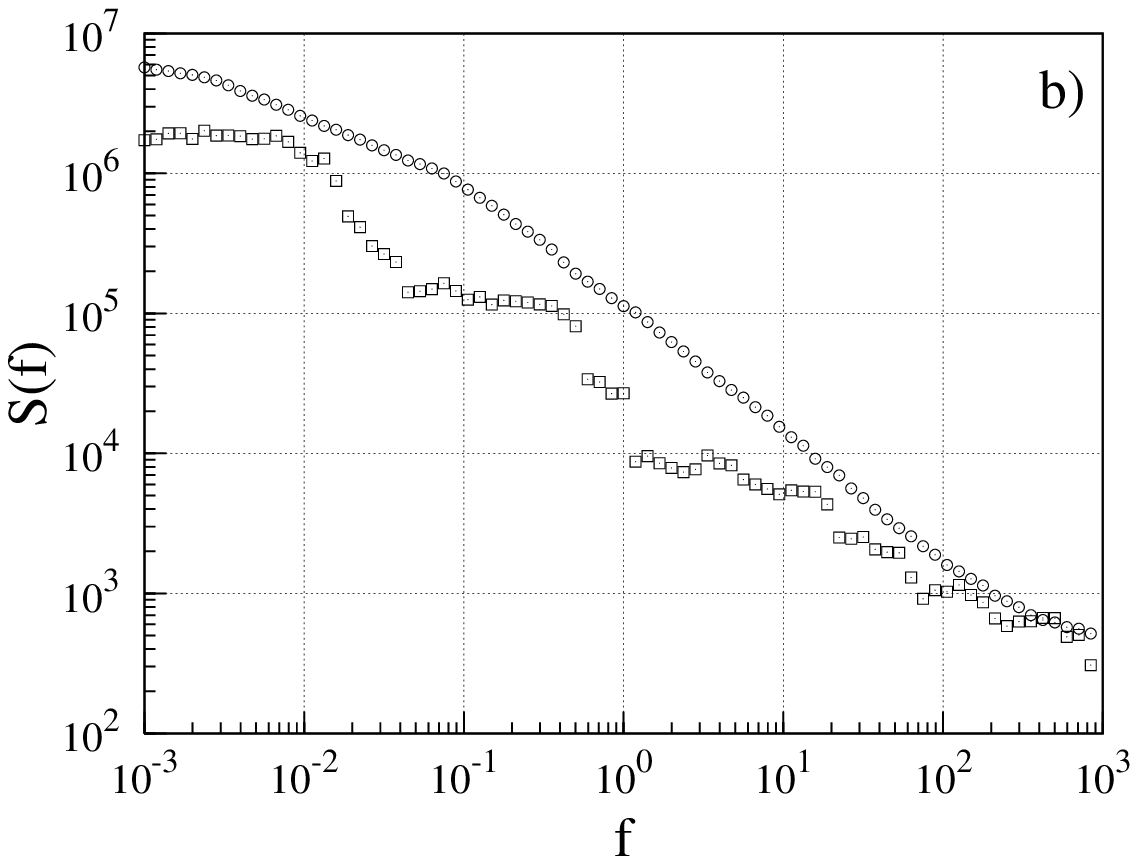}
\end{center}
\vspace{-10pt}
\caption{Distribution densities of the interpacket time $\tau_{k}$, (a), and the power spectra, (b), for the simulated 
point process (open circles) and the empirical data (open squares). The used parameters are as in the empirical data \cite{field,field2}, $\tau_{f}=0.101s, \tau_{k,min}=11.6 \mu{s}$ and $\tau_{k,max}=1000$~$\tau_{k,min}$}. 
\label{tit:dist}
\end{figure}

The files are requested from different servers located at different distance. This results in the distribution of the average inter-packet time $\tau_{p}$ in some interval. For reproduction of the empirical distribution of the interpacket time $\tau_{k}$ we assume the uniform distribution of $ \lg \tau_{k}$ in some interval $\left[ \tau_{k,min}, \tau_{k,max}\right]$, similarly to the McWhorter model of $ 1/f $ noise \cite{KG}. As a result, the presented model reproduces sufficiently well the observable non-Poissonian distribution of the arrival interpacket times and the power spectral density, as well (see Fig.~\ref{tit:dist}).    

\section{Conclusion}

In the paper it was shown that the processes exhibiting $1/f$ noise and the power-law distribution of the intensity may be generated starting from the signals as sequences of constant area pulses with the correlated appearance times as well as of different size Poissonian pulses. Combination of both approaches enables the modeling of the signals in the Internet traffic. 

\section*{Acknowledgment}
 
The support by the Lithuanian State Science and Studies Foundation is acknowledge.

\end{document}